\documentclass[prl,twocolumn,floatfix,showpacs,preprintnumbers,amsmath,amssymb,superscriptaddress]{revtex4}
\usepackage[T1]{fontenc}% pour utiliser les caract�res fran�ais
\usepackage{graphicx}% Include figure files
\usepackage{dcolumn}% Align table columns on decimal point
\usepackage{bm}% bold math
\begin{document}
%----------------------------------------------------------------------
%\title{Universality Classes in Fracture Roughness Scaling: a case study on planar cracks}
\title{Fracture Roughness Scaling: a case study on planar cracks} %{\bf( a shorter title )}}
%----------------------------------------------------------------------

\author{St{\'e}phane Santucci}
%\email{stephane.santucci@ens-lyon.fr}

\affiliation{Laboratoire de physique, UMR CNRS 5672, Ecole Normale Sup{\'e}rieure de Lyon, 46 all{\'e}e d'Italie, 69007 Lyon, France}
\affiliation{Physics of Geological Processes, Oslo University, 1048 Blindern, N-0316 Oslo, Norway}
\affiliation{Fysisk Institutt, Universitetet i Oslo, 1048 Blindern, N-0316 Oslo, Norway}

\author{M{\'e}lanie Grob}
\affiliation{Institut de Physique du Globe de Strasbourg, UMR CNRS 7516, EOST / Universit\'e de Strasbourg, 5 rue Ren{\'e} Descartes, F-67084 Strasbourg Cedex, France}

\author{Alex Hansen}
\affiliation{Institutt for fysikk, Norges teknisk-naturvitenskapelige Universitet, N--7491 Trondheim, Norway}

\author{Renaud Toussaint}
\affiliation{Institut de Physique du Globe de Strasbourg, UMR CNRS 7516, EOST / Universit\'e de Strasbourg, 5 rue Ren{\'e} Descartes, F-67084 Strasbourg Cedex, France}

\author{Jean Schmittbuhl}
\affiliation{Institut de Physique du Globe de Strasbourg, UMR CNRS 7516, EOST / Universit\'e de Strasbourg, 5 rue Ren{\'e} Descartes, F-67084 Strasbourg Cedex, France}

\author{Knut J{\o}rgen M{\aa}l{\o}y}
\affiliation{Fysisk Institutt, Universitetet i Oslo, 1048 Blindern, N-0316 Oslo, Norway}

%----------------------------------------------------------------------
\begin{abstract}
Using a multi-resolution technique, we analyze large in-plane fracture
fronts moving slowly between two sintered Plexiglas plates. We find
that the roughness of the front exhibits two distinct regimes
separated by a crossover length scale $\delta^*$.  
Below $\delta^*$, we observe a multi-affine regime and the measured roughness exponent $\zeta_{\parallel}^{-} = 0.60\pm 0.05$ is in   
agreement with the coalescence model.
Above $\delta^*$, the fronts are mono-affine, characterized
by a roughness exponent $\zeta_{\parallel}^{+} = 0.35\pm0.05$,
consistent with the fluctuating line model.
We relate the crossover length scale to fluctuations in fracture
  toughness and the stress intensity factor.
\end{abstract}
%----------------------------------------------------------------------
        %Interface structure and roughness
%----------------------------------------------------------------------
\date{\today}
\maketitle
%----------------------------------------------------------------------

Since the pioneering work of Mandelbrot et al.~\cite{mpp84}
demonstrating the self-affine character of fracture surfaces of
metals, numerous studies have been devoted to the morphology of
fracture surfaces \cite{b97,anz06}. In particular, the roughness
exponent $\zeta_{\perp}$ characterizing this self-affinity was shown
to be very robust and further on conjectured to be universal
\cite{blp90} with $\zeta_{\perp} \sim 0.8$ over a large set of
  materials and conditions \cite{blp90,anz06,mhhr92,b97} and up to very large scales \cite{sss95}. In the
  weak disorder limit when toughness fluctuations are small compared
  to stress loading fluctuations, there are data suggesting that
  $\zeta_{\perp}$ takes on a smaller value, 0.4 \cite{bppbg06,pavh06}.
  %or possibly show a smooth transition to a flat surface for no disorder.  
  A first attempt at investigating the origin of a
universal fracture roughness exponent in the quasistatic
  propagation limit was made by Hansen et al.~\cite{hhr91} who
suggested that in two dimensions it might be related to the directed
polymer problem.  This idea was further developed by R{\"a}is{\"a}nen
et al.~\cite{rsad98}: the fracture surface follows the surface that
minimizes the integrated strength of the intact material.  Numerical
studies based on this idea gave a roughness exponent
$\zeta_{\perp}=0.41\pm 0.02$.  A different idea was proposed by
Bouchaud et al.~\cite{bblp93}.  In their picture, the fracture surface
is the "footprint" of a passing fluctuating elastic line --- the
fracture front --- moving through a disordered three-dimensional
landscape.  This powerful idea opened up for the existence of {\it
  two\/} roughness exponents: one describing the roughness orthogonal
to the average crack plane $\zeta_{\perp}$ and another describing
the roughness of the front in the average roughness plane
$\zeta_{\parallel}$ \cite{dbl95}. 

In order to simplify the 3-d configuration, Schmittbuhl et al.\ \cite{srvm95}
proposed a numerical model where the crack front is constrained to propagate along a weak plane (suppressing the
  out-of-plane roughness). The in-plane roughness exponent was found equal to  $\zeta_{\parallel}=0.35$, and
%For planar cracks where the
 % fracture propagates along a weak plane, (suppressing the
 % out-of-plane roughness), Schmittbuhl et al.\ \cite{srvm95}
  %constructed a numerical model giving $\zeta_{\parallel}=0.35$.
  later on, refined to
  $\zeta_{\parallel}=0.39$ by Rosso and Krauth \cite{rk02}.  
  Even though such a fluctuating line approach 
  \cite{bsp08} can match several scaling exponents related to the crack front dynamics \cite{msst06}, 
  it fails at reproducing the in-plane roughness exponent $\zeta_{\parallel}$ measured up to
  now around 0.6 \cite{sm97,dsm99,ms01}.  We will adress precisely this problem in the present Letter, insisting on the fact that the fluctuating line model   %Actually, there are two aspects of the fracture process that it does not contain: first, it
  is typically a perturbative approach assuming the local slope of the
  front to be small, and ignoring crack coalescence
  \cite{bbfrr02}. A theory based on a mapping of the fracture process
  to a correlated percolation one \cite{hs03,shb03} considered
  precisely the latter aspect.  Numerical simulations based on this
  model gave $\zeta_{\parallel}=0.6$, substantially larger than the
  value found based on the fluctuating line model but consistent with the 
  experimental results obtained up to now.  The coalescence model also clarified the
  controversy over the concept of self-affinity
  \cite{az04,shb04,hbrs07}.

%The experimental situation has evolved markedly since universality was
%first proposed in 1990 \cite{blp90}. For an early review, see \cite{b97}.  
%We summarize the picture today as follows: in the strong disorder
%limit ({\it i.e.\/} when toughness fluctuations are of the order of stress 
%loading fluctuations), $\zeta_{\perp}$ takes the universal value 0.8. 
%This holds over a large set of materials and conditions \cite{blp90,anz06,mhhr92,b97}.
%In the weak disorder limit when toughness 
%fluctuations are small compared to stress loading fluctuations, there are
%data  suggesting that $\zeta_{\perp}$ takes on a smaller value, 0.4 
%\cite{bppbg06,pavh06} with a possible transition to a flat front for no 
%disorder. On the other hand $\zeta_{\parallel}$ has been
%measured to be about 0.6 \cite{sm97,dsm99,ms01}.  This is in good agreement
%with the coalescence model \cite{hs03,shb03} and in contradiction with the
%fluctuating line approach \cite{srvm95}.
%However, the fluctuating line model \cite{bsp08}
%matches other scaling exponents related to crackling noise measured
%experimentally \cite{msst06}.
%
%----------------------------------------------------------------------
\begin{figure}
\includegraphics[width=9 cm, height=4cm,clip]{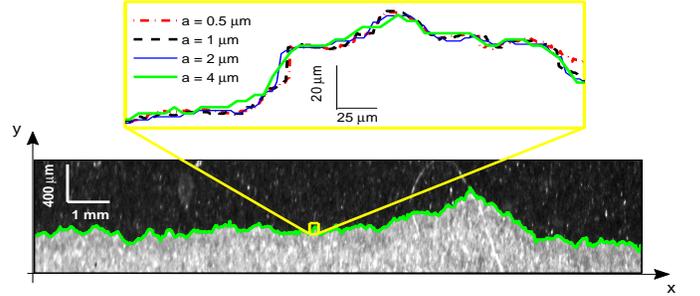}
\caption{A typical fracture (using glass beads of a diameter $\varnothing_1 \sim 50 \mu m$ 
  for roughening the samples)  and a zoom of the crack front
  $y(x)$  to emphasize the effect
  of the optical resolution given by the pixel size $a$.} 
\label{fig1}
\end{figure}
%----------------------------------------------------------------------

The goal of the present Letter is to address the possible
  coexistence of two roughness scaling regimes in the case of in-plane
  fracture.  We analyze stable mode I fracture fronts propagating
along the sand blasted and sintered contact plane between two PMMA
plates \cite{sm97,dsm99,ms01,msst06,mmhsvdbr07}. An important
contribution to our analysis comes from the compilation  of
  numerous observations at different resolutions. 
 We observe that the planar cracks follow two distinct scaling regimes, multi-affine at small scales and mono-affine at larger ones, characterized by different roughness exponent $\zeta^-_{\parallel}=0.60$ and  $\zeta^+_{\parallel}=0.35$, respectively.  
We show that those two regimes are separated by a
well-defined crossover length $\delta^*$ controlled by a balance between the
stress intensity factor variability along the fracture front and
fluctuations in the fracture toughness. Below  $\delta^*$, the value of the roughness exponent is in agreement with the prediction of the coalescence model
$\zeta_{\parallel}=0.60$,  while the large-scale
exponent is consistent with the value predicted by the fluctuating
line model, $\zeta_{\parallel}=0.39$.  We argue in the second part of
the manuscript why {\it both\/} theories may be correct in describing
the experiment, but operating at different length scales.

{\it Experiments --}The experimental setup allows for a stable mode I
crack propagation along a weak plane in a PMMA block from a
displacement imposed normally \cite{sm97}.  Toughness fluctuations
along the weak plane are artificially introduced during sample
preparation, which consists of annealing two sandblasted PMMA plates.
In order to modify the toughness fluctuations, we changed the type and
the size of the blasting particles using glass beads of diameters
around $\varnothing_1 \sim 50 \mu$m, $\varnothing_2 \sim 200 \mu$m, and $\varnothing_3 \sim 300 \mu$m
and a glass-aluminum powder with a typical particle size around $S
\sim 50 \mu m$.  We also changed the loading speed and procedure, with
interfaces recorded during their propagation at various velocities.
%, $0.4 \mu m /s < v < 40 \mu m /s$ or crack front at rest $v=0$.  
In order to obtain a multi-high resolution description of the fronts,
we considered fracture fronts at rest.  During those
experiments, after a slow crack propagation, the sample was unloaded
in order to arrest the crack.  Then, we took high resolution pictures
($3871\times 2592$ pixels) of the front at rest (Fig.\ \ref{fig1})
using a digital camera mounted on an optical microscope.  Using a
translation stage that can move the microscope in the $x$ direction
parallel to the front (and perpendicular to the fracture propagation
direction $y$) neighboring pictures were taken.  Up to 15 high
resolution pictures were then assembled resulting in fracture fronts
with around 25 000 data points and a pixel size $a = 0.48$ $\mu$m. To
remove acquisition artifacts, different resolutions of the front
description were obtained by changing the magnification of the optical
zooms (see Fig.~\ref{fig1}).  This results in images of the same
fracture at resolutions, 4, 2, 1 and $0.48$ $\mu$m per pixel with
respectively around 4000, 8000, 16000 and 25000 data points per image.
This procedure was repeated 20 times in order to obtain 20 independent
fracture fronts.  The actual total length of each analyzed crack was
around $15$ mm.
%----------------------------------------------------------------------
\begin{figure}
      \includegraphics[width=8.5 cm]{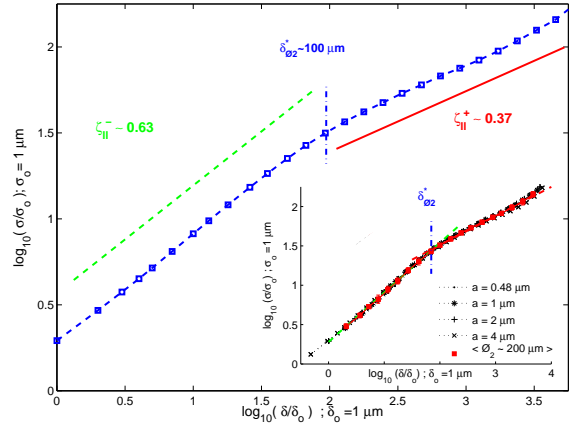} 
\caption{ Scaling behavior of the rms $\sigma$ of the height fluctuations  $\Delta y(\delta)$ 
with two different roughness exponents  $\zeta_{\parallel}^{-} \sim 0.63 $ and $\zeta_{\parallel}^{+} \sim 0.37 $, 
below and above the typical scale $\delta^* \sim 100$ $\mu$m respectively.  The inset shows that this scaling
is independent of the image resolution $a$. 
%We observe two different scaling behaviors of the root mean
% square (rms) $\sigma$ of the distributions $P[\Delta' y]$ where
% $\Delta'y(\delta) = \Delta y(\delta)- \langle \Delta y \rangle$,
%  below and above the typical scale $\delta^* \sim 100$ $\mu$m with
% respectively two different roughness exponents
%  $\zeta_{\parallel}^{-} \sim 0.63 $ and $\zeta_{\parallel}^{+} \sim
% 0.37 $. The filled square symbols correspond to the rms $\sigma$ of
% the distributions $P[\Delta' y]$ shown in inset: we draw in a
% semi-log plot, $P[\Delta' y(\delta)].\sqrt{2 \pi \sigma^2}$
%  vs.\ $[\Delta' y(\delta)]/ \sqrt{2 \sigma^2}$ for increasing length
%  scales $\delta$, shifted vertically for visual clarity. Above $\delta^*$, the
% dotted lines $f(x)=e^{-x^2}$ fit the parabolic shape of the Gaussian distributions. 
}\label{f2}
\end{figure}
%----------------------------------------------------------------------

{\it Two scaling regimes --} 
We analyze  %We base our analysis on the statistical distribution function of 
the height fluctuations of the crack fronts
$\Delta y(\delta) = y(x+\delta)-y(x)$ where $y(x)$ is the advance of
the front along the $y$ direction at position $x$. 
%Here, the experimental data set is large enough to perform a direct measurement
% of the pdf.  
We will first consider fracture fronts at rest for
samples prepared with $200$ $\mu m$ glass beads. Then, we will prove
the universality of our results showing the same analysis for various
experimental conditions.  In Fig.~\ref{f2}, we examine the scaling
behavior of the root mean square (rms) of the height fluctuations  $\sigma(\delta) =\langle \Delta
y^2(\delta)\rangle^{1/2}$.  At small scales below $\delta^* \sim 100$
$\mu$m, we observe a self-affine scaling behavior:
$\sigma(\delta)\propto \delta^{\zeta_{\parallel}^{-}}$ with a
roughness exponent $\zeta_{\parallel}^{-} = 0.60 \pm 0.05$. This is
consistent with previous experimental measurements \cite{sm97, dsm99,
  ms01, msst06} and the value predicted by the
coalescence model \cite{shb03}.  However, at scales larger than
$\delta^*$, we observe a crossover to another scaling regime with a
smaller roughness exponent $\zeta_{\parallel}^{+} = 0.35  \pm 0.05$. 
%It is important to remark that this is the first time that such a scaling
% regime is reported experimentally.  In terms of numerical modeling,
This value corresponds to the fluctuating line model prediction
$\zeta_{\parallel} = 0.39$ \cite{rk02}. 
Due to the limited scaling range for $\delta> \delta^*$, 
our data do not rule out a possible slow crossover to a flat front (no disorder regime), at large
scales \cite{Katzav07}.
In the inset of Fig.~\ref{f2}, we demonstrate the robustness of our results by
showing that the two different scaling regimes and the cross-over
length scale observed are independent of the sampling resolution of
the interface.  

\begin{figure}
\includegraphics[width=8.5 cm]{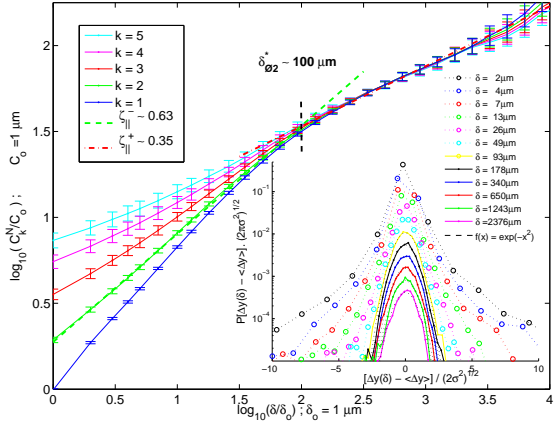}
\caption{ 
We observe two different scaling regime of the normalized structure functions $C_k^N(\delta) = C_k(\delta)/ R_k^G(\delta)$
separated by the cross-over $\delta^* \sim 100$ $\mu$m. 
The corresponding p.d.f. of the height fluctuations $\Delta y(\delta)$ are shown in inset where we plot in semi-log  $P[\Delta' y(\delta)].\sqrt{2 \pi \sigma^2}$
 vs.\ $[\Delta' y(\delta)]/ \sqrt{2 \sigma^2}$ for increasing length
  scales $\delta$, shifted vertically for visual clarity. Above $\delta^*$, the
  lines $f(x)=e^{-x^2}$ fit the parabolic shape of the Gaussians.}
\label{f3}
\end{figure}

In order to study in more details those different scaling behaviors, we develop a multi-scaling analysis \cite{mmhsvdbr07}
by performing a direct measurement %(thanks to the large experimental data set) 
of the pdf    $P[\Delta' y]$   of the height 
fluctuations $\Delta'y(\delta) = \Delta y(\delta)- \langle \Delta y \rangle$ and computing their structure functions 
$C_k(\delta)=\langle \vert \Delta y(\delta)\vert^k\rangle_x^{1/k}$. %, thanks to our large experimental data set.
On Fig.~\ref{f3}, we show the distributions of the height fluctuations  $P[\Delta' y(\delta)]$ for logarithmically increasing length scales $\delta$.  
Above the characteristic length scale $\delta^* \sim 100$ $\mu$m, the shape of the distributions is Gaussian, while for smaller length scales, we observe long tails consistent with a non-Gaussian and multi-affine scaling. Indeed, we show that  the structures functions $ C_k(\delta)$  --when normalized by the set of values  $R_k^G = \sqrt{2} \left({\Gamma \left( {{k+1} \over 2} \right) / \sqrt{\pi}}\right)^{1/k} $ corresponding to the ratios $R_k^G  = C_k^G(\delta)/ C_2^G(\delta)$ obtained for a Gaussian and mono-affine signal (see \cite{mmhsvdbr07} for details)-- collapse and follow a self-affine scaling with a unique roughness exponent $\zeta_{\parallel}^{+} = 0.35  \pm 0.05$  corresponding to the elastic line prediction.
Below $\delta^*$, the fanning of the structure functions  confirms the deviation to the Gaussian statistics and reveals an effective multi-affine behavior consequence of the heterogeneities along the interface leading to steep crack front slopes. %with a k-dependent roughness exponent. 
A fit to $C_2(\delta)$ on that range gives $\zeta_{\parallel}^{-} = 0.60 \pm 0.05$, in agreement with the coalescence model \cite{shb03}.
%----------------------------------------------------------------------
%----------------------------------------------------------------------
 
{\it Cross-over length scale -- }We now investigate what controls the
crossover length $\delta^*$.  We base our discussion on the Griffith
criterion that assumes a balance between the stress intensity factor
$K$ and the fracture toughness, $K_c$.  We introduce a mean
field argument to describe the stress intensity factor variation in
the direction of the propagation around the average position of the
front $\bar{y}$: $K(y)=K_0(\bar{y})+K'(y-\bar{y})$ where $K'=\partial
K /\partial y$ is the average local gradient of the stress intensity
factor.  Then, the toughness of the asperities along the weak plane
is supposed to be random around an average $K_c^*$ and uncorrelated
beyond the asperity size $\delta_c$. We assume that 
  $K_0(\bar y) = K_c^*$ and that the fluctuation of the toughness
over the front width $\sigma(\delta_c)$ reads as:  $K_c(\bar y
  \pm\sigma(\delta_c))=K_c^* \pm \Delta K_c (\delta_c)$  where
$\Delta K_c$ is the magnitude of the toughness fluctuations on scale
equal or larger than $\delta_c$. Finally, we estimate the width of the
crack front $\sigma(\delta_c)$ to be the typical scale in the
$y$-direction at which the failure criterion is met:
$K(\bar{y}+\sigma(\delta_c))\approx
K_c(\bar{y}+\sigma(\delta_c))$. Hence at a first order, we get an
estimate of the front width at the asperity scale as a function of the
magnitude of the toughness fluctuations $\Delta K_c$ and the local
stress gradient $K'$: $\sigma(\delta_c)= \Delta K_c / K'$. Due to the self-affinity of the front with a  roughness exponent
$\zeta_{\parallel}$, our argument leads to an estimate of the
prefactor of this scaling as:
\begin{equation}
\sigma(\delta)=
\sigma(\delta_c)\left(\frac{\delta}{\delta_c}\right)^{\zeta_{\parallel}}
= \left( \frac{\Delta K_c}{
  K'}\right)\left(\frac{\delta}{\delta_c}\right)^{\zeta_{\parallel}}.
\end{equation}
An important consequence is that the scaling of the
fracture front will be hidden in the no disorder limit: when either
the toughness disorder disappears ($\Delta K_c \rightarrow 0$) or when
the loading gradient becomes very large ($K' \rightarrow \infty$).

\begin{figure}[ht]
\includegraphics[width=9cm]{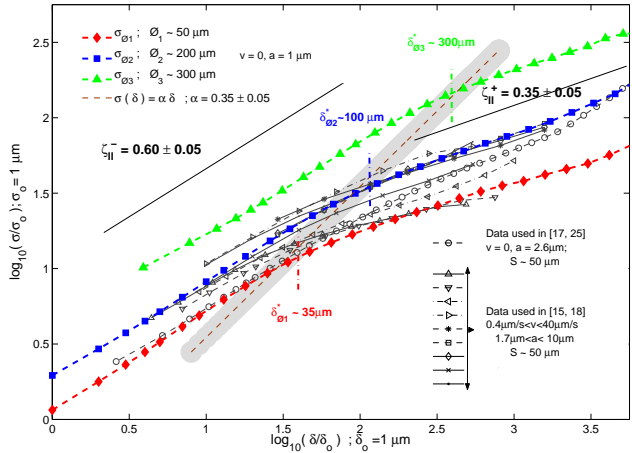}
\caption{Effect of disorder on the scaling of interfacial crack
  fronts.  We plot the rms $\sigma$ of the height fluctuations $\Delta
  y$ as a function of the scale $\delta$ for crack fronts at rest and
  samples blasted with glass beads of various diameters
  $\varnothing_1 \sim 50\ \mu$m, $\varnothing_2 \sim
  200\ \mu$m and $\varnothing_3 \sim
  300\ \mu$m. The line $\sigma(\delta)=0.35\ \delta$ separates the two
  scaling regimes. To insist on the robustness of our results, we add
  various data obtained during previous experiments with many
  different experimental conditions: glass-aluminium powder with a
  typical size around $S \sim 50 \mu m$ and crack front propagating at
  various velocities $v$ \cite{dsm99, ms01, msst06,
    mmhsvdbr07}. 
  }
\label{f4}
\end{figure}

We now address the estimate of the local slope of the crack front at
the asperity scale, {\it i.e.,\/} $\sigma(\delta_c)/\delta_c$.  Two
cases emerge.  First, if the local slope is small,
$\sigma(\delta_c)\ll \delta_c$ the front may be described using a
perturbative approach.  We expect in this case that the fluctuating
line model to be valid, leading to a roughness exponent
$\zeta_{\parallel}^+ \approx 0.39$.  We note here that
$\sigma(\delta)\ll \delta$ is valid for all $\delta > \delta_c$ if it
is fulfilled for $\delta_c$ due to the self affinity of the front.
The second situation occurs when $\sigma(\delta_c) \ge \delta_c$. In
this case, we assume the coalescence model to be valid with a
roughness exponent $\zeta_{\parallel}^- = 0.6$.  Hence, the slope at a
scale $\delta$ scales as $\sigma(\delta)/\delta \propto
\delta^{\zeta_{\parallel}^--1}$, which means that it decreases with
increasing $\delta$.  This implies that there is a scale $\delta^*$ at
which the slope $\sigma(\delta^*)/\delta^*=\alpha$ with $\alpha < 1$,
 and where the fluctuating line model is assumed to take
  over.  Subsequently, we estimate
\begin{equation}
\delta^* =\left( \frac{\Delta K_c}{\alpha
  K'}\right)^{1/(1-\zeta_{\parallel}^-)}
\delta_c^{-\zeta_{\parallel}^{-}/(1-\zeta_{\parallel}^{-})} .
\end{equation}
This length scale $\delta^*$ is different to the Larkin length
\cite{Blattter94} separating various pinning regimes \cite{Tanguy04}
of an elastic line in a random medium. It rather corresponds to the
onset of steep front slopes or overhangs (Fig.1) leading to deviations to the Gaussian and mono-affine scaling of the fronts (Fig.3),  and therefore 
limiting the range of validity of the elastic line models.
In Fig.~\ref{f4} we plot $\sigma(\delta^*)/\delta^*=\alpha$ for
$\alpha =0.35$ and check that it accounts for the crossover between
the two scaling regimes for the various experiments performed in
different conditions.  We conclude that the crossover length scale
$\delta^*$ is a function of the asperity size $\delta_c$, the
toughness fluctuations $\Delta K_c$ and the stress intensity factor
gradient $K'$.  Following the above arguments, the measure of the
crossover $\delta^*$ provides an estimate of the link between the
magnitude of the toughness fluctuations $\Delta K_c$ and the asperity
size $\delta_c$ knowing the loading conditions $ K'$.  
The crossover $\delta^*$ depends on
$\Delta K_c$ and $K'$ through a power law with a positive exponent
$1/(1-\zeta_{\parallel}^-)\approx 5/2$, very different from its
variation with $\delta_c$,
$-\zeta_{\parallel}^{-}/(1-\zeta_{\parallel}^{-})\approx -3/2$.  Also
the crossover $\delta^*$ is not expected to scale linearly with the
asperity size $\delta_c$ as suggested in \cite{Dalmas08} except if the
toughness fluctuations are proportional to the asperity size: $\Delta
K_c \propto \delta_c$.

{\it Disorder effect -- } In order to check the effect of disorder and material microstructure
({\it i.e.} $\Delta K_c$ and $\delta_c$), we modified the
heterogeneities of the sintered interface between the two Plexiglas
plates by preparing different samples using glass beads of different
diameters $\varnothing_1 \sim 50\ \mu$m, $\varnothing_2 \sim
200\ \mu$m and $\varnothing_3 \sim
  300\ \mu$m. % for the blasting procedure. 
%----------------------------------------------------------------------
We show in Fig.\ \ref{f4} the scaling behaviour of the interfacial
crack fronts with those various types of disorder that influence the
toughness fluctuations (unfortunately the measurement of the link
between sand-blasting particle size and toughness fluctuations was not
possible).  We observe mainly the same features as in previous figures
with the two different scaling behaviours separated by a
characteristic size respectively $\delta_{\varnothing_1}^*$, $\delta_{\varnothing_2}^*$ and $\delta_{\varnothing_3}^*$. 
For instance, when using
smaller glass beads ($\varnothing_1 \sim 50\ \mu$m), the amplitude of the height fluctuations of the
fronts decreases (vertical shift) as well as the scaling range at
small scales providing a roughness exponent $\zeta_{\parallel}^{-}
\sim 0.6 $ up to the scale $\delta_{\varnothing_1}^* \sim 35\ \mu$m.
We checked %the robustness of our results showing 
that all observations are independent of the sampling resolution, and
the analysis techniques  (see Fig.\ \ref{f2} and Fig.\ \ref{f3}).  Moreover, we also verified on Fig.\ \ref{f4}
that those results are consistent with the morphology of planar cracks
obtained during previous experiments \cite{dsm99, ms01, msst06,
  mmhsvdbr07} with various conditions concerning both the sample
preparation (glass beads mixed with an aluminum powder with a
wider size distribution) and the fracturing process with both crack
front at rest ($v=0$) or  propagating
at various velocities ($0.4 \mu m /s < v < 40 \mu m /s$).

{\it Conclusion -- } We have  analyzed the scaling
properties of long planar crack fronts moving along a 
rough interface between two sintered Plexiglas plates.  We identified two
scaling regimes separated by a length scale $\delta^*$ that depends on
the ratio of the local stress drop and the local toughness
disorder. Above $\delta^*$, the fronts are mono-affine, characterized
by a roughness $\zeta_{\parallel}^{+} = 0.35\pm0.05$,
consistent with the fluctuating line model.  Below $\delta^*$, we see
a different scaling regime, multi-affine, with a roughness exponent
$\zeta_{\parallel}^{-} = 0.60 \pm 0.05$. The later roughness
  exponent is in agreement with the coalescence model.  A similar
picture may explain observations for the scaling of crack surfaces
\cite{Daguier97, bppbg06, Dalmas08, Morel08} suggesting that crack
coalescence is the mechanism operating at small scales corresponding
to the process zone while the fluctuations of the elastic front line
is the dominating one at larger scales.

%----------------------------------------------------------------------
\begin{acknowledgments}
We thank D.\ Bonamy, E.\ Bouchaud, J.\ Mathiesen, S.\ Roux,
L.\ Vanel, M. Adda-Bedia, C. Marli\`ere, S. Zapperi, M. Alava and L. Laurson for helpful discussions.
\end{acknowledgments}
%----------------------------------------------------------------------

%----------------------------------------------------------------------
\end{document}